\documentclass[preprint,12pt]{elsarticle}

\usepackage{amssymb}
\usepackage{amsthm}

\usepackage{amsmath}  

\usepackage{xcolor}

\usepackage{hyperref}

\journal{Journal of Computational Physics}

\begin{document}

\begin{frontmatter}



\title{\bf Signal modelling with \\ Overdetermined Morphing Technique}


\author[1]{Nikita Belyaev\corref{cor1}}

\author[2,3]{Rostislav Konoplich}

\author[4]{Kirill Prokofiev}

\cortext[cor1]{Corresponding author}

\affiliation[1]{organization={Laboratory for Nuclear Science , Massachusetts Institute of Technology},
            addressline={77 Massachusetts Avenue}, 
            city={Cambridge},
            postcode={02139}, 
            state={MA},
            country={United States of America}}

\affiliation[2]{organization={Department of Physics, New York University},
            addressline={726 Broadway}, 
            city={New York},
            postcode={10003}, 
            state={NY},
            country={United States of America}}
            
\affiliation[3]{organization={Department of Mathematics and Physics, Manhattan University},
            addressline={4513 Manhattan College Parkway}, 
            city={Riverdale},
            postcode={10471}, 
            state={NY},
            country={United States of America}}
            
\affiliation[4]{organization={Department of Physics, the Hong Kong University of Science and Technology},
           addressline={Clear Water  Bay}, 
           city={Kowloon}, 
            state={Hong Kong},
            country={SAR China}}

\begin{abstract}
Precise modelling of a signal in processes with multiple observables, exhibiting a complex dependency on the underlying parameters, is often a difficult and challenging task. Predicting the results of experimental measurements in high-energy physics reactions serves a good example. The reaction rates and distributions of momenta of the final state particles, depend on the parameters of the underlying physics model in a non-trivial way. The conventional way to predict the experimental of observables is to use the Monte Carlo morphing technique on a finite discrete set of simulated samples.  In this article we extend this technique by using the overdetermined morphing basis. We show that our approach yields a better result than the traditional technique in terms of the statistical power and uniformity of the resulting prediction, and delivers better precision. We further demonstrate that the overdetermined morphing is better suited for description of extended regions of phase space than the conventional morphing. We use the modelling of kinematic distributions of the final state particles produced in decays of non-Standard Model Higgs bosons as an example.
\end{abstract}

\end{frontmatter}


\section{Morphing Technique}
\label{sec:morph}
The experimentally measured observables in particle physics experiment usually consist of event rates and kinematic distributions of the final state particles.
Fundamental physics parameters are usually extracted from the experimental observables using a likelihood fit to a pre-defined physics model. The conventional method of modelling the experimental observables consists of Monte Carlo generation of high-energy physics reaction events for a particular set of model parameters, followed by the simulation of the detector response. This method is not readily applicable to study continuous ranges of model parameters: in order to predict the behaviour of experimental observables as a function of continuous model parameters, an extrapolation between basis Monte Carlo samples is required. In modern particle physics experiment this  problem is usually solved by using some variation of the so-called sample morphing technique~\cite{Baak}, \cite{Brenner2016}, \cite{Kyle}, \cite{Verkerke}. Before introducing the improvements, which are the main subject of this paper, we shall briefly review the main principles of morphing.

The following assumptions lie in the core of the morphing technique: 
\begin{itemize}
 \item{The modelled process can be described with a quantum mechanical matrix element, which has a polynomial dependence on its couplings;}
 \item{The experimental observables can be modelled with the square of the matrix element.}
\end{itemize}
It can be noted that in principle this approach is not limited to quantum processes, but can be applied to any process, which can be modelled with a polynomial characteristic function. We define experimental observable $T(\alpha)$, where $\alpha=(\alpha _1 .. \alpha _i)$ is a vector of the fundamental parameters defining physics model. Its dependence on the parameters of the model can be then expressed as follows:
\begin{align}\label{eq:obs}
    T(\alpha)&\sim |M(\alpha _1 .. \alpha _i)|^2 \\ \nonumber
   & = \alpha _1 ^2 |O_1|^2 + .. + \alpha _i ^2 |O_i|^2  \\
  & +2\alpha _1\alpha _2\Re (O_1^*O_2) +.. +2\alpha _{i-1} \alpha _i \Re (O_{i-1}O_i). \nonumber
\end{align}
Here $M(\alpha _1 .. \alpha _i)$ represents the quantum mechanical matrix element, and $O_1 .. O_i$ are its parts, corresponding to individual fundamental parameters $\alpha _ 1 .. \alpha _i$. The experimental observable 
$T(\alpha)$ can now be modelled as a linear sum of independently generated Monte
Carlo samples. The samples corresponding to $|O_i|^2$ can be generated setting $\alpha _i$ to unity and all other $\alpha$ parameters to zero: $|O_i|^2 \sim |M(\alpha _i =1, \alpha _{k\neq i} =0)|^2$. The samples, corresponding to the mixed  terms of type $\Re (O_i^* O_j)$ can be produced by generating a sample with only $\alpha _i$ and $\alpha _j$ parameters activated, while all others are set to zero, and  by subtracting the corresponding $|O_i|^2$ and $|O_j|^2$, such that:
\begin{align*}
    \Re (O_i^* O_j) \sim &|M(\alpha _{i,j} =1, \alpha _{k\neq i,j} =0)|^2 \\
    -& |M(\alpha _i =1, \alpha _{k\neq i} =0)|^2  - 
    |M(\alpha _j =1, \alpha _{k\neq j} =0)|^2.
\end{align*}
An experimental observable, corresponding to an arbitrary set of fundamental parameters $\alpha _{\rm target} = (\alpha _1 .. \alpha _n)$ can now be modelled using observables obtained from a finite set of $N$ Monte Carlo samples generated for a fixed vectors of values of fundamental constants $\alpha ^i _{\rm input}$:
\begin{equation}\label{eq:MorphFunc1}
   T(\alpha _{\rm target})=\sum ^{N}_i w_i(\alpha _{\rm target}, \alpha ^i _{\rm input}) T_i (\alpha ^i _{\rm input}),
\end{equation}
where the set of weight coefficients $w_i(\alpha _{\rm target}, \alpha ^i _{\rm input})$ is known as {\it the morphing function}.  Expanding the Eq.~\ref{eq:MorphFunc1} to the level of individual fundamental constants one gets the expressions for the individual weight coefficients:
\begin{align}\label{eq:m_mat}
T(\alpha _{\rm target})&=\\
&\nonumber \\
&(a_{11}\alpha _1^2 + a_{12}\alpha _2^2 +..+a_{1,n+1}\alpha_1\alpha _2+..+a_{1,N}\alpha _{N-1}\alpha _N ) T_1(\alpha ^{1} _{\rm input})\nonumber \\ \nonumber
+&(a_{21}\alpha _1^2 + a_{22}\alpha _2^2 +..+ a_{2,n+1}\alpha_1\alpha _2+..+ a_{2,N}\alpha_{N-1}\alpha_N) T_2(\alpha ^{2} _{\rm input})\\ \nonumber
+&..\\ \nonumber
+&(a_{N1}\alpha _1^2 + a_{N2}\alpha _2^2 +..+ a_{N,n+1}\alpha_1\alpha _2+..+ a_{N,N}\alpha_{N-1}\alpha_N) T_N(\alpha ^{N} _{\rm input}).
\end{align}
The task of modelling $T(\alpha _{\rm target})$ using the set of input samples $T_i (\alpha ^i _{\rm input})$ is then reduced to finding $N^2$ coefficients $a_{ij}$ in Eq.~\eqref{eq:m_mat}. The latter can be achieved by requiring $T(\alpha _{\rm target})= T_i (\alpha ^i _{\rm input})$, formulating the problem as a system of linear equations $A\Lambda = I $, or:
\begin{align}\label{eq:lambda_def}
    &{\left( 
    \begin{tabular}{cccc}
    $a_{11}$& $a_{12}$&..&$a_{1N}$\\
    $a_{21}$& $a_{22}$&..&$a_{2N}$\\
      & &..&\\ 
      $a_{N1}$& $a_{N2}$&..&$a_{NN}$\\
    \end{tabular}
    \right)\cdot} &  \\ 
    &&\nonumber\\
   &{ \left(
    \begin{tabular}{cccc}
    $(\alpha ^1 _{{\rm input},1})^2$&$(\alpha ^2 _{{\rm input},1})^2$&..&$(\alpha ^N _{{\rm input},1})^2$\\
    $(\alpha ^1 _{{\rm input},2})^2$&$(\alpha ^2 _{{\rm input},2})^2$&..&$(\alpha ^N _{{\rm input},2})^2$\\    
      & &..&\\ 
     $\alpha ^{1} _{{\rm input},n}\alpha ^1 _{{\rm input},n-1}$ & $\alpha ^{2} _{{\rm input},n}\alpha ^2 _{{\rm input},n-1}$ &..& $ \alpha ^{N} _{{\rm input},n}\alpha ^N _{{\rm input},n-1}$\\
    \end{tabular}
     \right )} & = I_{N} \nonumber, 
\end{align}
where $I_N$ is the $N \times N$ identity matrix. 
The unique morphing function for modelling $T(\alpha _{\rm target})$ using a set of $N$ input samples $ T_i (\alpha ^i _{\rm input} )$ can be found by inverting the square matrix $\Lambda$ of rank $N$. Each column in matrix $\Lambda$ represents all possible independent combinations of powers of input parameters. 

A common example in high-energy physics is modelling a process with one production and one decay vertices. An experimental observable is then proportional to a matrix element:
\begin{equation}
T(\alpha)\sim |M(\alpha _1 .. \alpha _i)|^2=\left|\left(\sum _{i=1} ^{p+s} \alpha _i O_i^{\mu} \right) \Pi_{\mu \nu} \left( \sum _{j=1} ^{d+s} \alpha _j O_j^{\nu}\right)\right|^2,
\end{equation} 
where $\Pi_{\mu \nu}$ is a propagator that is independent of couplings. Symbols $p$ and $d$ denote the numbers of independent couplings to be taken into account in the production and decay vertices respectively. The symbol $s$ denotes the number of couplings shared by the production and decay vertices.
The number of simulated samples $N$ required for modelling such a two-vertex process can be obtained using the following relation:
\begin{align}\label{eq:n_sam}
N &= \frac{1}{24}s(s+1)(s+2)\left [ (s+3) + 4 (p+d)\right ]
\nonumber \\
&+\frac{1}{4} \left [ s(s+1)p(p+1) 
+ s(s+1)d(d+1)+p(p+1)d(d+1)\right ] \\ 
&+\frac{1}{2}pds (p+d+s+3).
\nonumber
\end{align}

It should be noted that all the considerations discussed in this section can be applied to processes having a higher degree polynomial dependence than shown in Eq.~\ref{eq:obs}. The structure of the resulting matrix $\Lambda$ will be more complex, but all the main conclusions will remain.
\section{The quality of the morphing basis \label{sec:qmb}}
A set of $N$ samples $ T_i (\alpha ^i _{\rm input} )$ required for modelling a process with $n$ fundamental parameters is known as the {\it morphing basis}. The optimal selection of input fundamental constants $\alpha ^i _{\rm input}$ represents the main challenge of the morphing approach.  The input constants should be selected such, that the resulting matrix $\Lambda$ is invertible and the corresponding solution is stable. At the same time, the selected morphing basis should result in a good statistical description of the target region of parameter space. The stability of the solution can be ensured by selecting the input fundamental constants such, that the resulting matrix $\Lambda$ is well-conditioned. For a nonsingular matrix the condition number \cite{Horn2012} with respect to 1-norm is giving by
\begin{equation}
\kappa(\Lambda) \equiv cond(\Lambda) = \|\Lambda \|_1\cdot \|\Lambda ^{-1}\|_1,\,
\end{equation}
where in terms of the matrix elements
\begin{equation*}
\|\Lambda \|_1 = 
\max\limits_{1\leq j\leq N} \sum ^N _{j=1} |\lambda _{ij}|.
\end{equation*}
By definition, $\kappa(\Lambda)$ is always greater than 1. A small $\kappa(\Lambda)$ indicates that the problem is well-conditioned, while a large $\kappa(\Lambda)$ suggests that the problem is ill-conditioned and it is possible to lose $log_{10}\kappa(\Lambda)$ digits of accuracy in solution, compared to precision 
of $\Lambda$. 

 To assess the quality of statistical description resulting from the morphing procedure in the current study we use the following two ratios:
\begin{equation}\label{eq:stat_norm}
\mathfrak{N}_1=\frac{\sum _i w_i \sigma _i}{\sum _i |w_i| \sigma _i} \;\;{\rm and}\;\; \mathfrak{N}_2=\frac{\left( \sum _i w_i \sigma _i \right)^2}{\sum _i (w_i \sigma _i)^2},
\end{equation}
where $\sigma _i$ are the individual cross sections of simulated input samples.
The effective weights ratio $\mathfrak{N}_1$ is defined as the ratio of the effective number of events resulting from the morphing procedure at a particular point of the  parameter space to the total number of events used as the morphing input. Since the morphing weight $w_i$ can be positive or negative, $0\leq\mathfrak{N}_1\leq 1$. The effective sample ratio $\mathfrak{N}_2$ represents the so-called Kish effective sample size \cite{Kish1965}, \cite{Kish}, indicating the number of events in the Monte Carlo simulated morphed sample having the same statistical power as a hypothetical simulated sample. In other words, Kish's formula $\mathfrak{N}_2$ is exactly the number of unit-weight Monte Carlo events that would deliver the same variance and hence the same statistical power as the morphed, weighted sample at the given parameter point.

\begin{figure}[htp]
  \centering
  \includegraphics[width=0.99\linewidth]{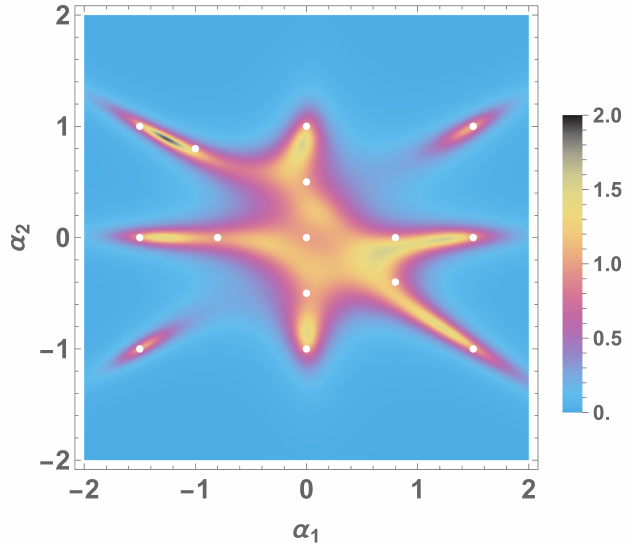}
  \caption{
  An example of distribution of the effective sample ratio $\mathfrak{N}_2$ as a function of two parameters $g_1$ and $g_2$ for a morphing basis of $15$ input samples.
  White circles indicate the position of the base samples in $(g_1,g_2)$ space. }
  \label{fig:nonuniform}
\end{figure}

There are several factors that reduce the precision of the modelling when using the conventional morphing technique. The main problem is that the corresponding morphing basis represents a fixed minimal set of input samples. The resulting model fails to uniformly describe the multidimensional parameter space due to cancellations that occur when different morphing terms are combined.  Consequently, the statistical power of the reconstructed distributions varies significantly from point to point, leading to significant uncertainties in certain regions. Fig. \ref{fig:nonuniform} shows an example of distribution of the norm $\mathfrak{N}_2$ as a function of two parameters $g_1$ and $g_2$ for a morphing basis of $15$ input samples.
The statistical power is rapidly deteriorating in the regions away from the base samples. The overall description is largely non-uniform. Since the size of the morphing basis is fixed by the number of parameters in the underlying model, there is little room for improvement. Adding an extra base sample to improve the statistical description in a problematic region would require an exclusion of one of the initial samples, rendering it redundant. In addition, morphing coefficients are calculated using simulated cross sections without accounting for the corresponding uncertainties. This results in substantial deviations between the morphed and simulated cross section values in the regions far from the base points. 

\section{Overdetermined systems of equations}
To address the challenges outlined in the previous section, we propose a new morphing technique that involves the least squares solution \cite{Strang2006} (QR-factorization) to an overdetermined system of equations, where the size of the morphing basis exceeds the number of independent combinations of powers of input parameters, determining the size of the matrix $\Lambda$ in Eq.~\eqref{eq:lambda_def}. To illustrate the underlying idea, let us first consider an example, featuring a general system of $M$ real-valued linear equations with $L$ real unknowns.  In the matrix form this system can be represented as:
\begin{equation}
    \mathbf{y} = B\mathbf{x}, \label{eq:system}
\end{equation}
where the $M\times L$ matrix $B$  and the vector $y$ are known, while the vector $x$ is to be found. If the number of equations is greater than the number of unknowns (i.e. $M>L$), this is the so-called overdetermined system of equations, which generally has no exact solution. The approximate solution of the system (\ref{eq:system}) can, however, be found by minimizing the residual vector $\mathbf{r}$:
\begin{equation}
    \mathbf{r} = \mathbf{y} - B\mathbf{x}, \label{eq:residual}
\end{equation}
using some norm. While any choice of norm is possible, the general preference is the 2-norm (sum of squared errors), which provides a closed-form solution to the approximation problem and is more convenient for numerical purposes \cite{Cadzow2002}. This approximate solution is usually referred to as the least-squares solution. The Eq.~\eqref{eq:residual} for the 2-norm case:
\begin{equation}
    \|\mathbf{r}\|_2 = \sqrt{(\mathbf{y} - B\mathbf{x})^2}, \label{eq:residual2}
\end{equation}
is minimized by calculating the gradient of this expression and setting this gradient to zero. The gradient with respect to $\mathbf{x}$ of $\|\mathbf{y} - B\mathbf{x}\|^2_2$ is:
\begin{equation}
    \nabla_{\mathbf{x}} \|\mathbf{y} - B\mathbf{x}\|^2_2 = \nabla_{\mathbf{x}} (\mathbf{y}^T\mathbf{y} - 2\mathbf{y}^T B\mathbf{x} + \mathbf{x}^T B^T B\mathbf{x}) = -2B^T \mathbf{y} + 2B^T B\mathbf{x} = 0.
\end{equation}

As long as matrix $B$ is full rank, the square matrix $B^T B$ of size $L \times L$ is positive definite, and the approximate solution of system \eqref{eq:residual2} is given by the system of normal equations:
\begin{equation}
    \mathbf{x} = (B^T B)^{-1} B^T \mathbf{y}.   \label{eq:normal}
\end{equation}
For this solution, the residual vector $\mathbf{r}$ is orthogonal to every vector in the span of $B$. The matrix $(B^T B)^{-1} B$ is the Moore-Penrose pseudo-inverse of the matrix $B$ \cite{Lawson1995}. 

In the case of an overdetermined system, the condition number of matrix $B$, the $\kappa(B)$, defined in Sec.~\ref{sec:qmb}, describes how well or how poorly the system of equations~\eqref{eq:system}, can be approximated.
For the matrix $B^T B$, the condition number is given by $\kappa(B^T B) = (\kappa(B))^2$. Since the condition number is by definition greater than unity, the solution \eqref{eq:normal} is significantly less well-defined than a solution to \eqref{eq:system}, involving only the matrix $B$. To avoid this loss of precision, an alternative approach should be employed. One possibility is to use the QR factorisation method. Here, the matrix  $B$ is expressed as $B = QR$, where $B$ is an $M \times L$ matrix, $Q$ is an $M \times L$ orthogonal matrix, which satisfies $Q^T Q = I$. The matrix $R$ is an $L \times L$ non-singular upper triangular matrix. 
The square of the matrix $B$ can be then replaced by $B^T B = R^T (Q^T Q) R = R^T R$, and the system \eqref{eq:normal} becomes:
\begin{equation*}
R^T R \mathbf{x} = R^T Q^T \mathbf{y}.
\end{equation*}
An approximate solution is then provided in the form:
\begin{equation}
      \mathbf{x} = R^{-1} Q^T \mathbf{y}.    \label{eq:qrsolution}
\end{equation}
Since $Q$ is orthogonal and does not change the condition number, the precision is maintained, as $\kappa(R) = \kappa(B)$. This  solution is numerically more stable against round-off errors compared to the normal equations. For well-defined matrices, both the least squares solution \eqref{eq:normal} and the QR solution \eqref{eq:qrsolution} yield almost identical results. 

The idea of overdetermined morphing consists of embedding the QR factorisation into the morphing technique to achieve better signal modelling in the regions of the phase space located far from the basis samples.

If the number of equations exceeds the number of unknowns (i.e., $M > N$), the matrices A and $\Lambda$ in 
Eq~\eqref{eq:lambda_def} should be expanded into non-square $N \times M$ and $M \times N$ matrices: 
\begin{align}\label{eq:lambda_def2}
    &{\left( 
    \begin{tabular}{ccccc}
    $a_{11}$ & $a_{12}$ & .. & $a_{1N}$ \\
    $a_{21}$ & $a_{22}$ & .. & $a_{2N}$ \\    
      &       & .. &      \\
    $a_{N1}$ & $a_{N2}$ & .. & $a_{NN}$ \\
      .. & .. & .. & .. \\
    $a_{M1}$ & $a_{M2}$ & .. & $a_{MN}$ \\
    \end{tabular}
    \right)} &  \\ 
    &&\nonumber\\
   &{\cdot \left(
    \begin{tabular}{cccccc}
    $(\alpha ^1 _{i,1})^2$ & $(\alpha ^2 _{i,1})^2$ & .. & $(\alpha ^N _{i,1})^2$ & .. & $(\alpha ^M _{i,1})^2$ \\
    $(\alpha ^1 _{i,2})^2$ & $(\alpha ^2 _{i,2})^2$ & .. & $(\alpha ^N _{i,2})^2$ & .. & $(\alpha ^M _{i,2})^2$ \\
      &                     & .. &                     & .. &                        \\
    $\alpha ^{1} _{i,n}\alpha ^1 _{i,n-1}$ & $\alpha ^{2} _{i,n}\alpha ^2 _{i,n-1}$ & .. & $ \alpha ^{N} _{i,n}\alpha ^N _{i,n-1}$ & .. & $ \alpha ^{M} _{i,n}\alpha ^M _{i,n-1}$ \\
      .. & .. & .. & .. & .. & .. \\
    $\alpha ^{1} _{i,m}\alpha ^1 _{i,m-1}$ & $\alpha ^{2} _{i,m}\alpha ^2 _{i,m-1}$ & .. & $ \alpha ^{N} _{i,m}\alpha ^N _{i,m-1}$ & .. & $ \alpha ^{M} _{i,m}\alpha ^M _{i,m-1}$ \\
    \end{tabular}
     \right )} & = I_{M} \nonumber, 
\end{align}

By performing a QR factorization of the $\Lambda$ matrix as $\Lambda = QR$, and then calculating $Q^T$ and inverting the square $N \times N$ matrix R, the approximate solution to the overdetermined system of equations can be obtained according to Eq~\eqref{eq:qrsolution}.   

\section{Physics example}
To illustrate the advantage of overdetermined morphing over the conventional technique, we apply both methods to model the production of a massive scalar Higgs boson $H$ in vector boson fusion (VBF) and its subsequent decay into a pair of vector bosons:
\begin{equation*}
VV\to H \to VV,
\end{equation*}
where $V=W,Z$.  
It is important to note that the morphing method is independent of the model's parametrization, requiring only a polynomial dependence of the quantity of interest on the parameters.

Assume there are two couplings in the production and decay vertices:  the standard model coupling, fixed at unity and one BSM coupling, that is variable and is the same in production and decay.  According to the morphing technique (Eq.~\ref{eq:n_sam}), the minimal number of base samples required to describe the model in the BSM coupling parameter space is $5$.  

\begin{figure}[ht]
  \centering
  \includegraphics[width=0.49\linewidth]{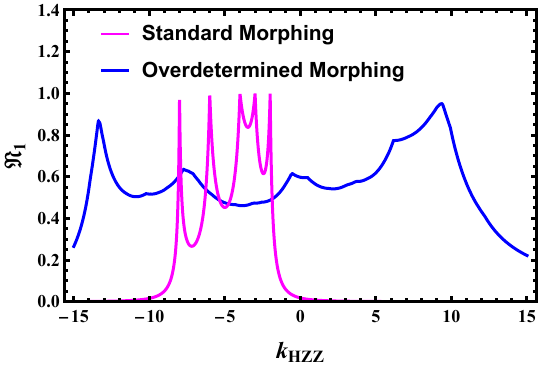}
  \hfill
  \includegraphics[width=0.49\linewidth]{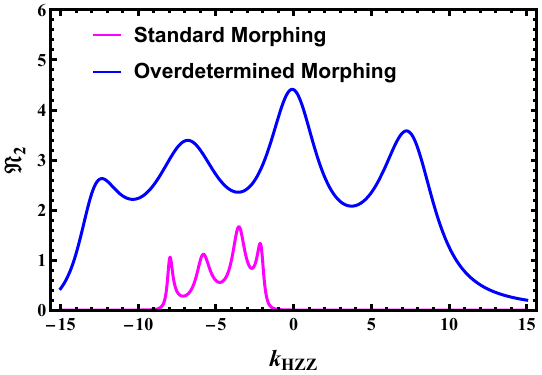}
  \caption{The effective weights ratio $\mathfrak{N}_1$ (left) and the effective sample ratio $\mathfrak{N}_2$ (right) as a function of the $kHZZ$ parameter value. The magenta curve corresponds to the standard morphing based on the minimal number of base samples with $kHZZ = (-8, -6, -4, -3, -2)$. The blue curve represents overdetermined morphing using $kHZZ = (-14, -12, -10, -8, -6, -4, -3, -2, 0, 1, 2, 4, 6, 8, 10)$.}
  \label{fig:Comparison_Morph_vs_Overdet}
\end{figure}

\begin{figure}[ht]
    \centering
    \includegraphics[width=0.99\linewidth]{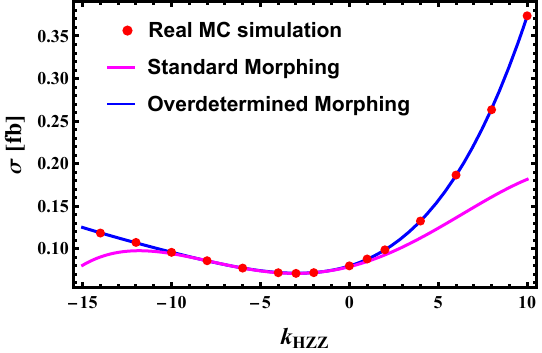}
    \caption{Cross section of the VBF process as a function of the $kHZZ$ parameter value. The magenta curve corresponds to the standard morphing for base samples $kHZZ = (-8, -6, -4, -3, -2)$. The blue curve gives the cross section obtained by the overdetermined morphing with $kHZZ = (-14, -12, -10, -8, -6, -4, -3, -2, 0, 1, 2, 4, 6, 8, 10)$. Red dots represent simulated cross sections.}
    \label{fig:xsec}
\end{figure}

To make an example of such configuration, we have used the MadGraph Monte Carlo generator within the framework of Higgs Characterization model~\cite{Artoisenet_2013}. The Standard Model Higgs boson couplings are fixed and the coefficient $kHZZ$, corresponding to a BSM CP-even operator is varied. The results of simulations are presented in Fig.~\ref{fig:Comparison_Morph_vs_Overdet} and Fig.~\ref{fig:xsec}. 
Fig.~\ref{fig:Comparison_Morph_vs_Overdet} shows the effective weights ratio $\mathfrak{N}_1$ and the effective sample ratio $\mathfrak{N}_2$ as functions of the $kHZZ$ coupling value.  The standard morphing exhibits highly non-uniform statistical power, with peaks corresponding to the base samples. In contrast, the overdetermined morphing technique provides a more uniform distribution, which can be extended to any region by adding additional samples. 
In the interval \(-8 \lesssim k_{HZZ} \lesssim -1\) the two strategies deliver comparable per–generated-event statistical power: the overdetermined morphing attains \(\mathfrak{N}_2\) values about three times larger than the standard scheme, but it is also built from three times more MC events, so the Kish efficiency \(\mathfrak{N}_2 / N_{\text{gen}}\) is similar. Outside that window, however, the standard morphing rapidly loses power (\(\mathfrak{N}_2 \ll 1\)), while the overdetermined construction maintains \(\mathfrak{N}_2 \gtrsim \mathcal{O}(1)\), thus providing a clearly dominant effective statistics across most of the parameter space. This illustrates that the overdeterminated approach largely removes the sharp ``basis-point'' instabilities visible with the minimal morphing.
 Note that the $\mathfrak{N}_1$ quantity is useful for verifying the correctness of the procedure in the case of the standard morphing: the peaks of the $\mathfrak{N}_1$ distribution should occur at coupling values corresponding to the base samples and reach a value of one.  

In Fig.~\ref{fig:xsec} the standard morphing correctly reproduces the cross sections near the base samples but can deviate significantly outside this region. Overdetermined morphing expands the region of valid cross sections by including additional samples.

Next, consider the process where the Higgs boson is  produced in gluon fusion (ggF) and decays into a pair of gauge bosons:
\begin{equation*}
gg\to H \to VV.
\end{equation*}
Simulation of this process was performed using the MadGraph generator within the framework of the Higgs Characterisation model~\cite{Artoisenet_2013}. We assume that in the production vertex there is only the Standard Model coupling and that it is fixed at unity. Three couplings are considered for the decay vertex: the standard model coupling (fixed at unity) and two couplings $k_{HZZ}$ and $k_{AZZ}$ that are variable and correspond to CP-even and CP-odd BSM terms of the model's Lagrangian. 

\begin{figure}[ht]
    \centering
    \includegraphics[width=0.49\linewidth]{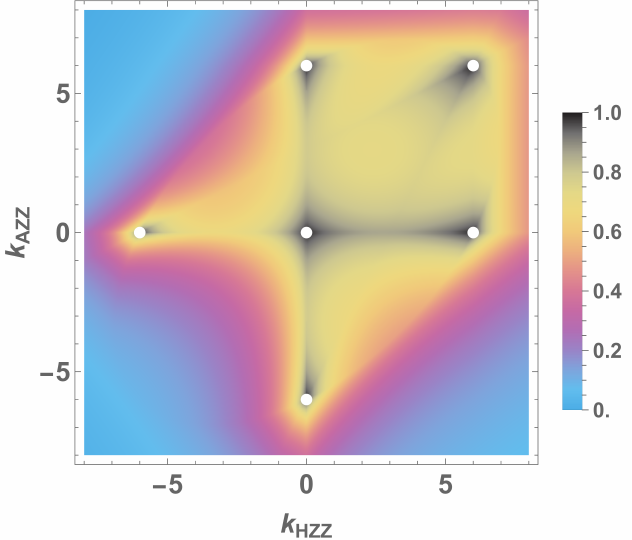}
    \hfill
    \includegraphics[width=0.49\linewidth]{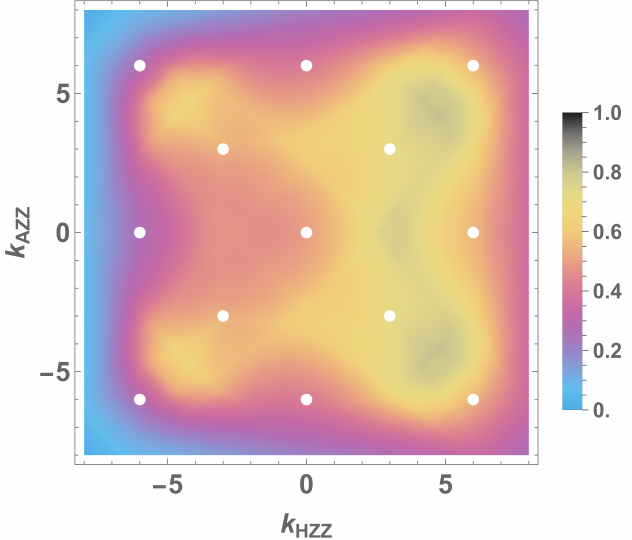}
     \caption{Density plot for the effeciive weights ratio $\mathfrak{N}_1$ vs parameters $k_{HZZ}$ and $k_{AZZ}$. White dots correspond to base samples.}
    \label{fig:neffntot}
\end{figure}

Fig.~\ref{fig:neffntot} shows the density plot for the effective weights ratio $\mathfrak{N}_1$. As expected, for the standard morphing it reaches maximum value of $1$ at the base samples, with statistical power dropping fast with the increase in ``distance" from the base points. The overdetermined morphing technique, however, significantly extends the region of acceptable statistics, reducing gaps that could complicate likelihood scans over the parameter space. The corresponding improvement is demonstrated on the right plot of Fig.~\ref{fig:neffntot}).

\begin{figure}[ht]
    \centering
    \includegraphics[width=0.49\linewidth]{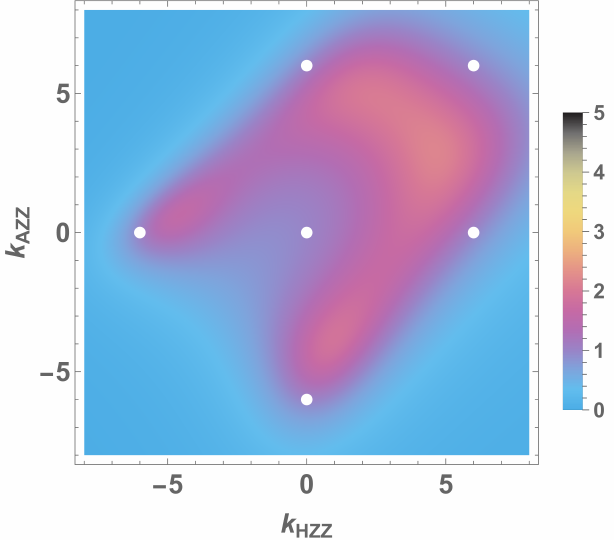}
    \hfill
    \includegraphics[width=0.49\linewidth]{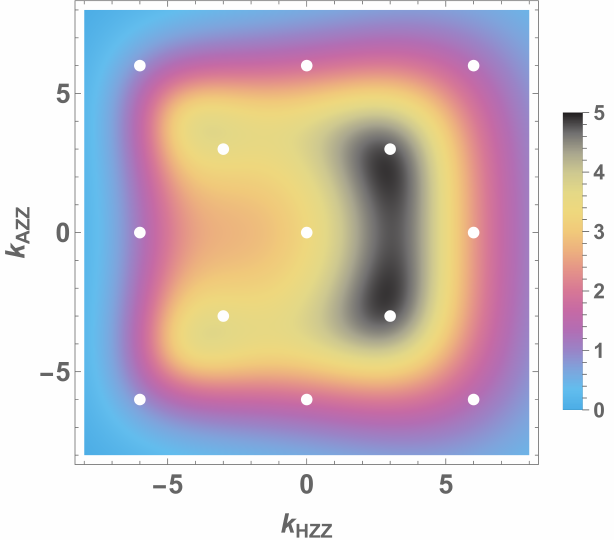}
    \caption{Density plot for effective sample ratio $\mathfrak{N}_2$ as a function of parameters $k_{HZZ}$ and $k_{AZZ}$. White dots correspond to base samples.}
    \label{fig:neffntot2}
\end{figure}

As discussed in Section~\ref{sec:morph}, the effective sample ratio $\mathfrak{N}_2$ represents the so-called Kish's effective sample size \cite{Kish}, indicating the number of Monte Carlo events in a simulated sample having the equal statistical power as the morphed sample. Fig.~\ref{fig:neffntot2} shows that while peak values per number of samples are similar for both types of morphing, the overdetermined morphing covers the larger area, leading to better precision.

\begin{figure}[ht]
    \centering
    \includegraphics[width=0.49\linewidth]{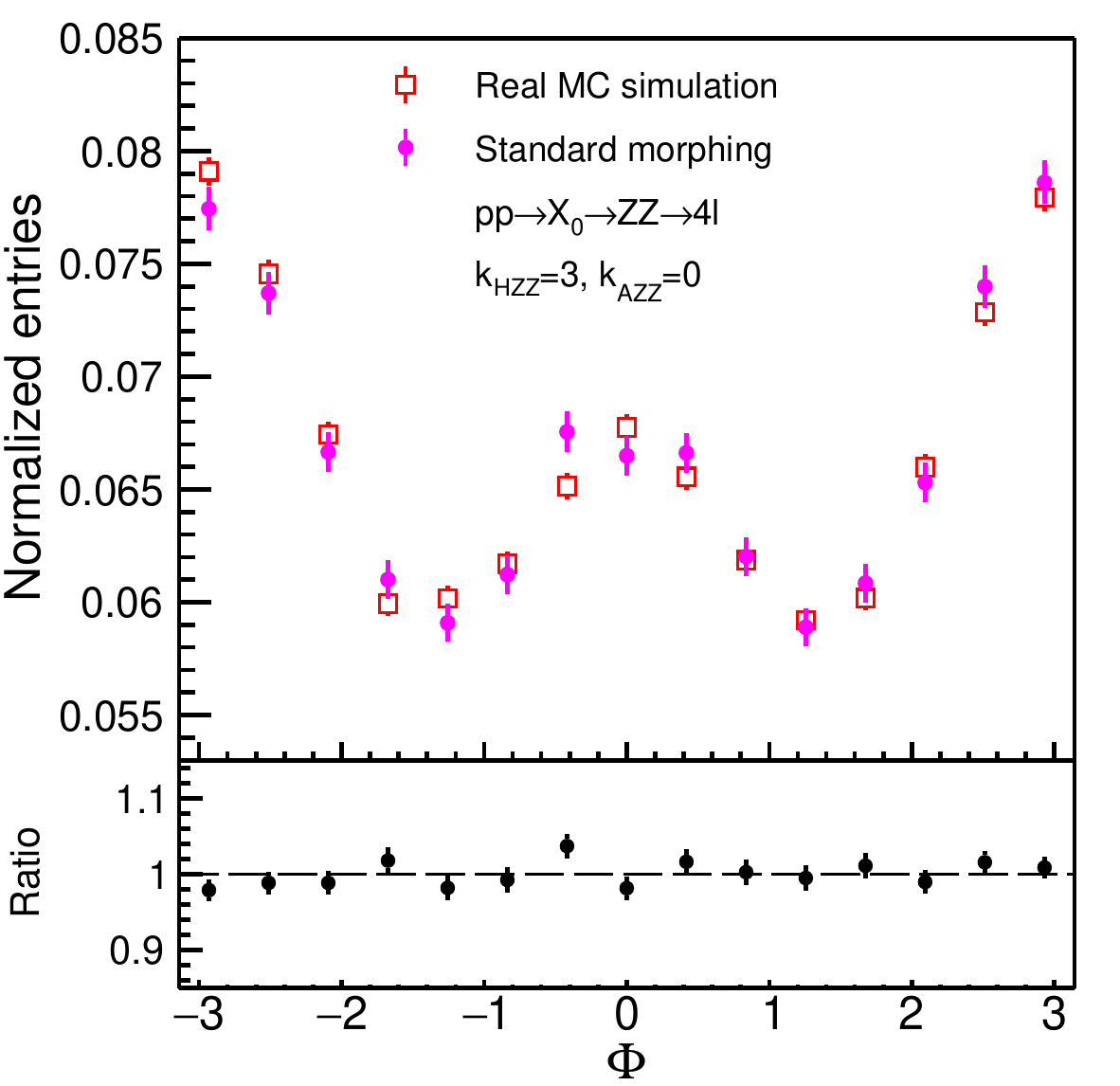}
    \hfill
    \includegraphics[width=0.49\linewidth]{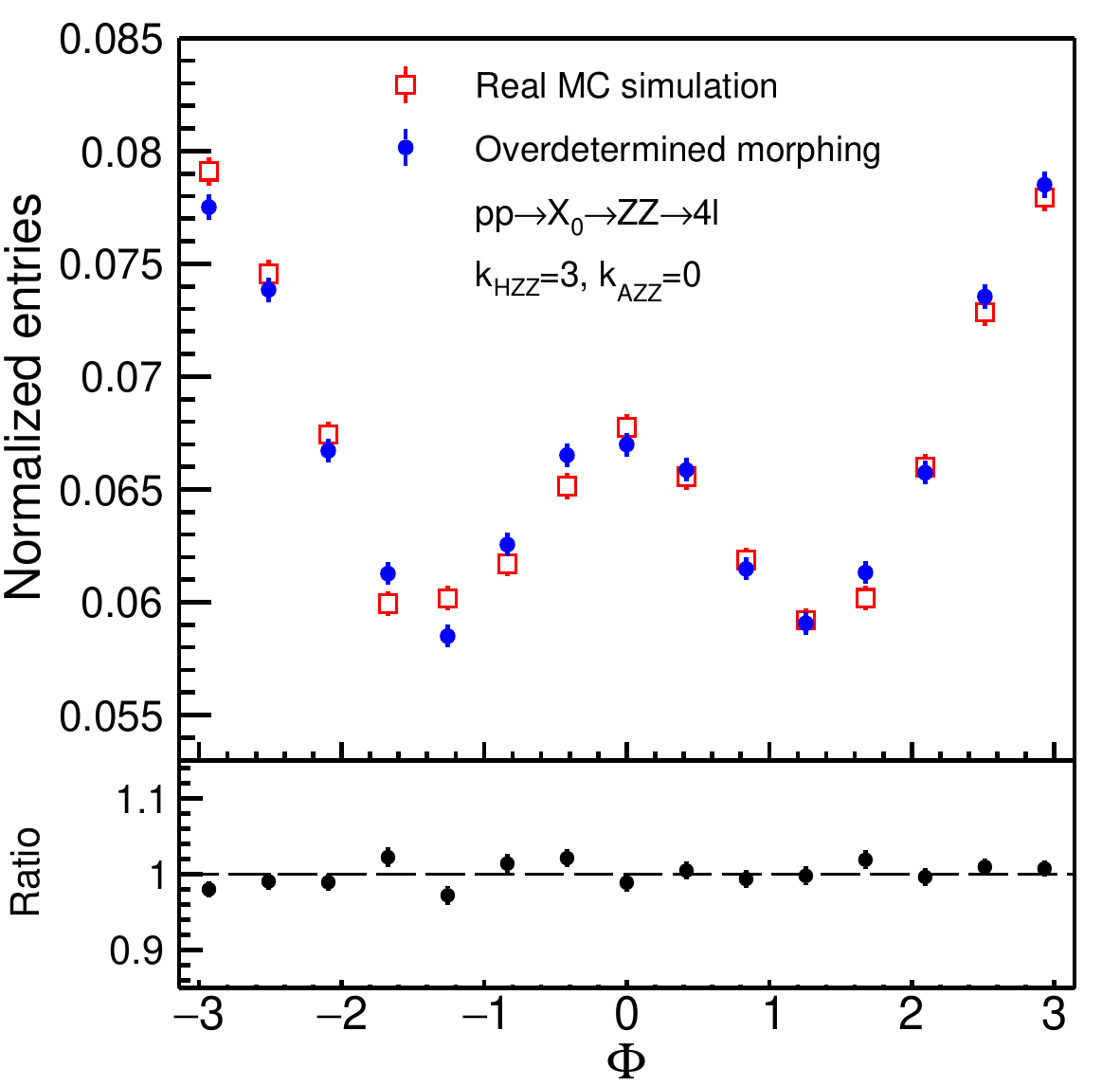}
    \caption{Normalized $\Phi$ - distribution for $k_{HZZ} = 3$ and $k_{AZZ} = 0$. Red points represent simulation results. Blue points represent the results of morphing, with the left plot showing the standard morphing and the right plot showing the overdetermined morphing.}
    \label{fig:Phi_3_0}
\end{figure}

\begin{figure}[ht]
    \centering
    \includegraphics[width=0.48\linewidth]{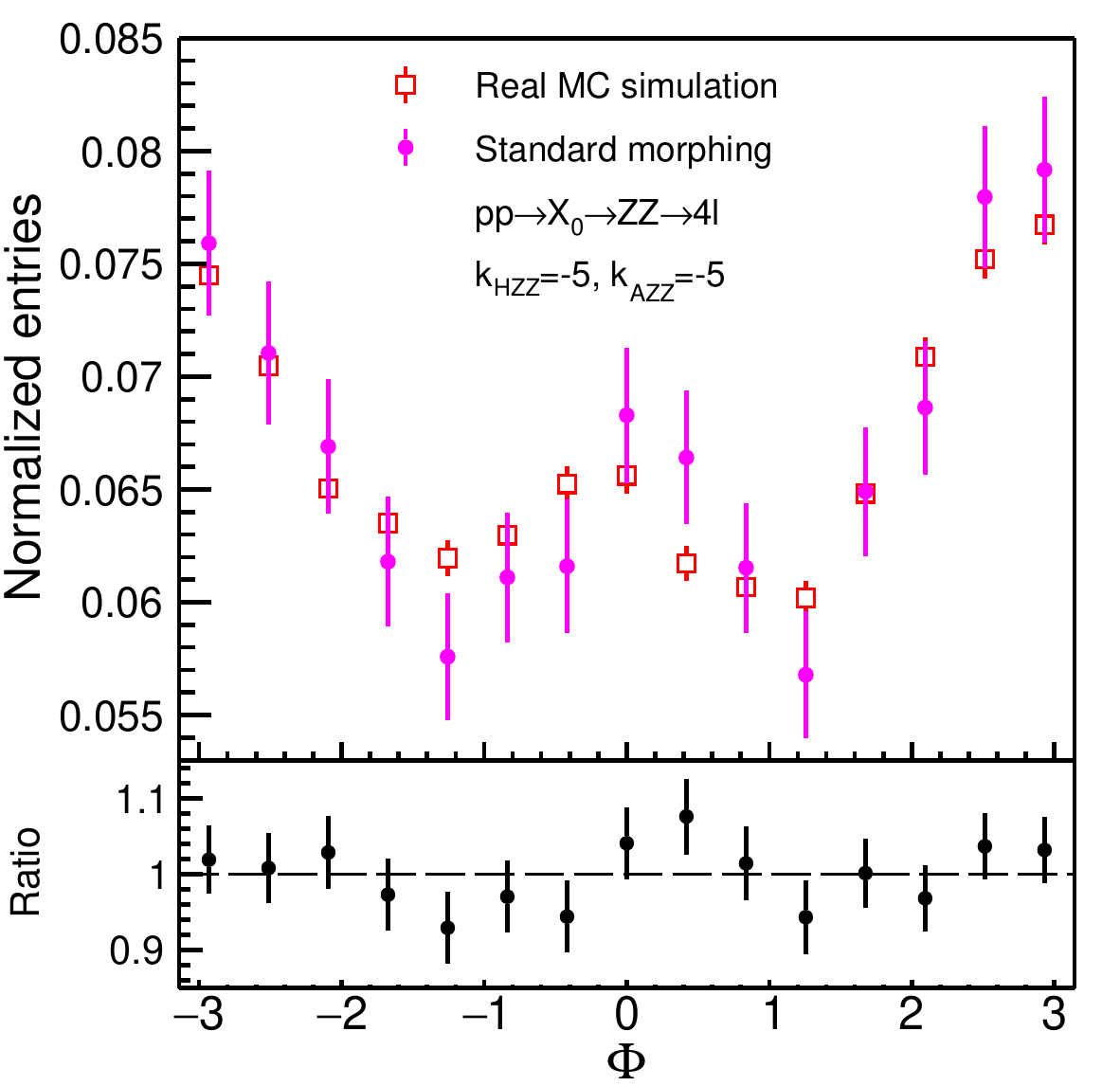}
    \hfill
    \includegraphics[width=0.48\linewidth]{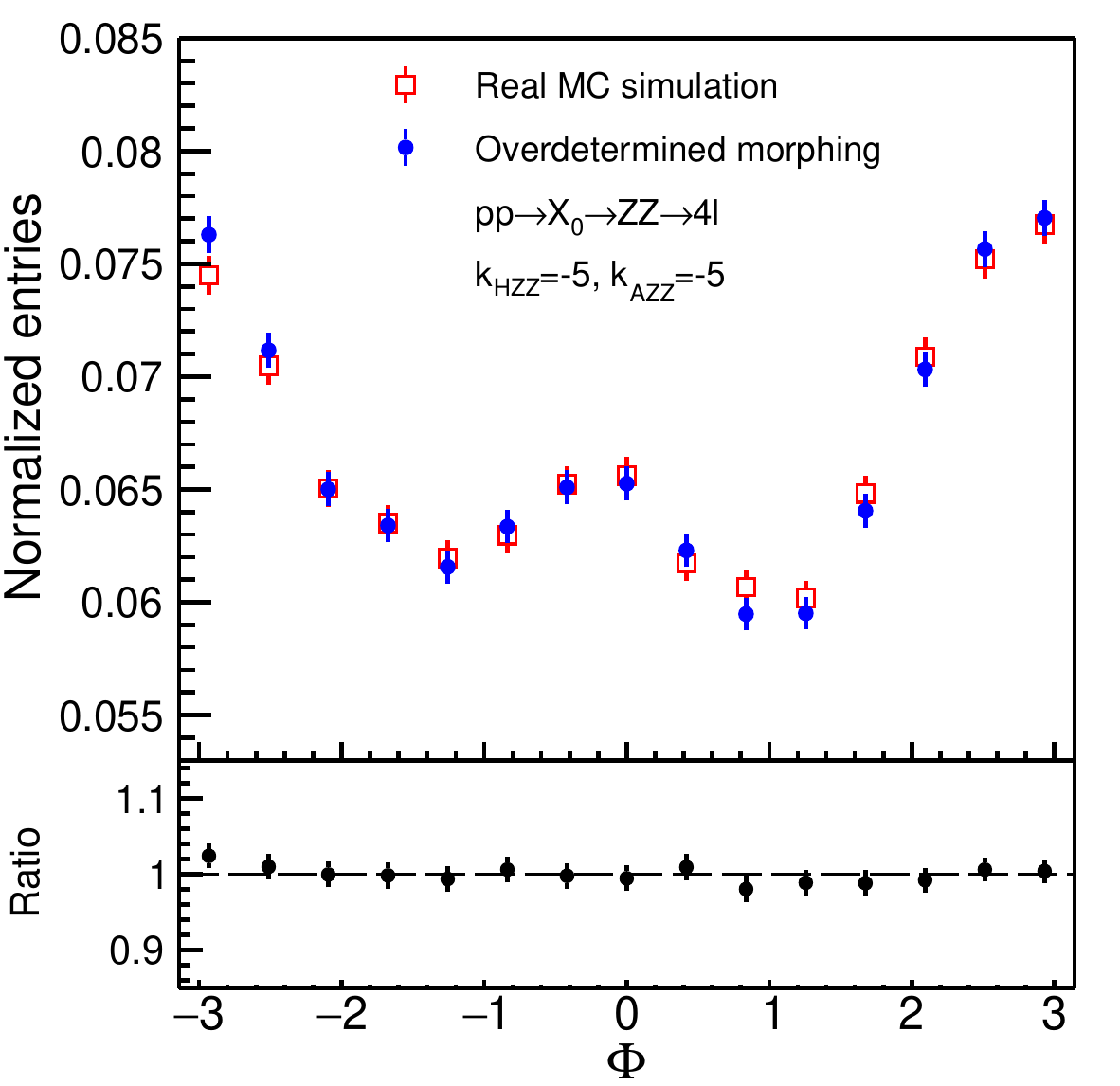}
    \caption{Normalized $\Phi$ - distribution for $k_{HZZ} = -5$ and $k_{AZZ} = -5$. Red points represent simulation results. Blue points represent the results of morphing, with the left plot showing the standard morphing and the right plot showing the overdetermined morphing.}
    \label{fig:Phi_m5_m5}
\end{figure}

Figs.~\ref{fig:Phi_3_0} and \ref{fig:Phi_m5_m5} display normalized \(\Phi\) - distributions for two test points, where \(\Phi\) is the azimuthal angle between the two decay planes spanned by the opposite-sign lepton pairs in the Higgs boson rest frame. For each base and test point, $50000$ events were generated. 
The red dots on the plots represent the simulated results with the corresponding uncertainties, while the blue points indicate the results of morphing of the base samples at the test points, also with uncertainties. For the test points $(k_{HZZ} = 3, k_{AZZ} = 0)$ and $(k_{HZZ} = -5, k_{AZZ} = -5)$, the overdetermined morphing is in good agreement with the simulated distributions, with comparable statistical uncertainties. Although the standard morphing provides almost similar results at the test point $(k_{HZZ} = 3, k_{AZZ} = 0)$, its uncertainties increase significantly at $(k_{HZZ} = -5, k_{AZZ} = -5)$, due to the reduced effective size number. 

To demonstrate the capabilities of the overdetermined morphing approach presented in this paper and to illustrate its practical applications, a standalone code was developed within Wolfram Mathematica software system~\cite{Mathematica}. This code can be found in a GitLab repository~\cite{Gitlab}. In this repository, the file ``OverdeterminedMorphing.wl" contains a working example of the Overdetermined morphing calculation.

\section{Summary}
In this study, we have identified limitations of the standard morphing technique and proposed an overdetermined morphing approach as a possible solution to these limitations. The new method addresses one of the most important issues of classical morphing: the strict limitation on the number of base samples, which results in unsatisfactory statistical performance in large regions of the phase space. This limitation arises from the requirement for the classical morphing matrix to be invertible, thus strictly defining the number of base samples for every problem at hand. The proposed new approach uses the overdetermined system of equations, allowing to employ a number of base samples, greater than the one dictated by the classical approach. Due to its ability to insert additional samples into the problematic regions, the proposed new method provides a more uniform distribution of statistical power across a broader parameter space, compared to the standard morphing. Moreover, the quality of the statistical description in arbitrary phase-space regions of interest can be brought to the desired level by inserting additional base samples. The power of the new approach is demonstrated on the modelling of the processes involving Higgs boson production and decay.

\section{Acknowledgements}
The work of RK was partially supported by the Kakos Endowed Chair in Science Fellowship.

 \bibliographystyle{elsarticle-num} 
 \bibliography{cas-refs}





\end{document}